\documentclass[twocolumn,showpacs,preprintnumbers,amsmath,amssymb]{revtex4}

\usepackage{graphicx}
\usepackage{dcolumn}
\usepackage{bm}

\begin{document}

\title{ Temperature Dependence of the Dielectric Constant and Resistivity of Diluted Magnetic Semiconductors.}

\author{M.P. L\'opez-Sancho and L. Brey. }

\affiliation{\centerline{Instituto de Ciencia de Materiales de
Madrid (CSIC),~Cantoblanco,~28049,~Madrid,~Spain.}}

\begin{abstract}

We study the effect that the ferromagnetic order has on the
electrical properties of Diluted Magnetic Semiconductors. We
analyze the temperature dependence of the dielectric constant and
of the resistivity of Ga$_{1-x}$Mn$_x$As.  In our treatment the
electronic structure of the semiconductor is described by a six
band Kohn-Luttinger Hamiltonian,  the  thermal fluctuations of the
Mn magnetic moments are treated in the mean field approximation,
the carrier-carrier interaction within the random phase
approximation, and the transport properties using the relaxation
time approximation.  We find that the Thomas-Fermi length
changes near 8$\%$ when going from the ferromagnetic to the
paramagnetic phase. We also find, in good agreement with the
experiments, that the resistivity changes near 20$\%$ when going
from zero  to the Curie temperature. We explain this change in the
resistivity in terms of the variation of the Fermi surface and the
transport scattering time when going from the ferromagnetic phase to
the paramagnetic phase.

\end{abstract}

\maketitle

The incorporation of Mn atoms into III-V semiconductors by
low-temperature MBE techniques was a big advance for the
integration of the spin degree of freedom in the semiconductor
technology. In Ga$_{1-x}$Mn$_x$As, and  for Mn concentrations
larger than $x \sim 1\% $, the magnetic ion  substitutes a cation,
introducing a $S=5/2$ local moment and a hole in the valence band
of the host semiconductor. These compounds present ferromagnetic
order, with a Curie temperature, $T_C$ near 100K
\cite{Matsukura,Matsukura-bis}, and this behavior is generally
called carrier induced ferromagnetism because the hole carriers
mediate the ferromagnetic coupling between the manganese
ions\cite{Dietl,Jungwirth,chatto,mjcdms}. The optimal Mn
concentration is near $x\sim 0.05$ and these materials are known
as Diluted Magnetic Semiconductors (DMS).
Experimentally\cite{Ohno} it is found that the presence of
compensating defects reduces the number of carriers in the system
being the density of carriers, $p$, much smaller than that of
magnetic ions, $c$. For the optimal Mn concentration and not very
large carriers densities, most of the magnetic properties of the
system seems to be reasonable well described by using a virtual
crystal approximation (VCA) \cite{Dietl,Jungwirth} that neglects
thermal and quantum fluctuations as well as disorder effects.
These effects are more important as the carrier density in the
system becomes larger\cite{Schliemann,mjcdms,dagotto-g}.
Post-growth annealing reduces the number of compensating defects,
increases the density of mobile holes in the semiconductor and
increases the value of the Curie
temperature\cite{Potashnik,Edmonds}. In the carrier induced
ferromagnetism model $T_C$ is proportional to the carrier density
of states at the Fermi energy, and for this reason the annealed
samples have a higher $T_C$ than the as-growth samples. Another
important difference between the annealed and the as-growth
samples is that the later ones present a metal-insulator
transition at $T_C$ that is absent in the former
ones\cite{Potashnik,Edmonds}. In the annealed DMS samples the
resistance increases with temperature ($T$) near 25$\%$ when going
from zero to $T_C$ and remains practically constant when  $T$ is
further increased.  As stated above, the post-growth annealing
process reduces the number of defects in the system and the
resistivity curves of the annealed samples reflect more
defect-free intrinsic properties of the diluted magnetic
semiconductor.

Most of the theoretical works on DMS study the origin of the
ferromagnetic order: the effects that the electronic properties of
the host semiconductor have on the ferromagnetic ground state and
the value of $T_C$. In this paper we study the $T$ dependence of
the electronic properties of the DMS. We analyze how   the spin
polarization of the holes affects the electronic properties of the
system and from that we obtain the temperature dependence of the
dielectric constant and of the conductivity of the doped
semiconductor.

The main result of our paper is presented in Fig.1. There, the $T$
dependence of the electrical resistivity for Ga$_{1-x}$Mn$_x$As
with $x \sim 0.05$ is plotted for two different hole densities.
The variation of the resistivity when going from zero $T$ to $T_C$
is near $20\%$, in good agreement with experiments in the more
intrinsic post growth annealed samples \cite{Potashnik,Edmonds}.
As we discuss below, this variation of the resistivity with $T$ is
due to the dependence of the Fermi surface and scattering times on
the  carrier spin polarization and therefore on $T$. The model we
present explains the observed  behavior of the electrical
resistivity with temperature.

In our calculations we describe the electronic structure  of the
DMS  using the six band Kohn-Luttinger Hamiltonian and the
disorder and thermal effects are treated in a mean field
approximation.  The dielectric constant is calculated in
the Random Phase Approximation (RPA)
formalism and the conductivities using the relaxation time
approximation.

The system is described by the following Hamiltonian,
\begin{equation}
H=H_{holes}+J\sum_{I,i} \mathbf{S} _I \cdot \mathbf{s}_i \, \,
\delta (\mathbf{r}_i -\mathbf{R}_I)  \, \, \, \,  , \label{Htotal}
\end{equation}
Where $H_{holes}$ is the part of the Hamiltonian which describes
the itinerant holes and the last term is the antiferromagnetic
exchange interaction between the spin of the Mn$^{2+}$ ions
located at $\mathbf{R}_I$ and the spin $\mathbf{s}_i$ of the
carriers. In order to obtain the temperature dependence of the
electronic  and magnetic properties of the system,
we minimize the free energy per unit volume,
\begin{equation} \label{Freeener}
\mathcal{F}= \mathcal{F}_{ions}+\mathcal{F}_{holes}\, \, \, \, \,
,
\end{equation}
where $\mathcal{F}_{ions}$ is the contribution of the ion spins to
the free energy and in the mean  field description has the form,
\begin{equation}
\mathcal{F}_{ions}= -T \, c \,  \log { \frac{ \sinh{(hS/T)}}
{\sinh{(h/2T)}}} \, \, \, \, \label{freeiones}
\end{equation}
being $h=Jp\xi/2$ and $\xi$ the spin polarization of the carriers.
$\mathcal{F}_{holes}$ is the free energy of the holes, which is
obtained in the virtual crystal approximation (VCA) using a
Luttinger $\mathbf{k}\cdot \mathbf{p}$ model for describing the
carriers. In the VCA the average density of states for the real
system is replaced by that of the average Hamiltonian. This approach
implies a translational invariant system with an effective
magnetic field acting on the carrier spins $H_{eff}=J S c m $,
being $m$ the polarization of the Mn spins. In  the Luttinger
$\mathbf{k}\cdot \mathbf{p}$ model the wave function of the holes
in the state $(n,\mathbf{k})$, where $n$ is the subband index and
$\mathbf{k}$ is the wave vector, is expressed as
\begin{equation}\label{wavefunction}
\psi_{n,\mathbf{k}} (\mathbf{r}) = e ^{i \mathbf{k} \, \mathbf{r
}} \sum_{J,m_j} \alpha_{n,\mathbf{k}} ^{J,m_J} |J,m_J> \, \, \, \,
,
\end{equation}
where $|J,m_J>$ are the six $\Gamma _{4v}$ valence band wave
functions. The coefficients $\alpha_{n,\mathbf{k}}^{J,m_J}$ and the
corresponding eigenvalues, $\varepsilon _ {n,\mathbf{k}}$, depend
on the spin polarization of the Mn, and are obtained, from the
Luttinger $\mathbf{k}\cdot \mathbf{p}$
Hamiltonian\cite{Dietl-bis,Albolfath}.

By minimizing Eq.(\ref{Freeener}) we obtain the $T$-dependence of
the spin polarization of the carriers, $\xi (T)$ and  of the Mn,
$m (T)$. In the VCA the Curie temperature  has the expression $
T_c = 2/3 \,  c J^{2} n_{\sigma}(\mu)$ where $ n_{\sigma}(\mu) $
is the density of states per site and spin at the Fermi
level\cite{Dietl-bis,Albolfath}. Through the dependence of the
Hamiltonian  on the spin polarization, we obtain the temperature
dependence of the eigenvectors, eigenvalues and chemical
potential. Along this paper we consider always the optimal Mn
concentration of $x$=0.05\cite{Ohno} and an exchange coupling
$J$=0.060$eVnm^3$\cite{Okabayashi}. Although the properties of the
system depend on the orientation of the
magnetization\cite{Dietl-bisAlbolfath}, this dependence is much
smaller than the T-dependence and since  we are interested in the
variation of the electronic properties with $T$  we fix the Mn
spin polarization in the $z$-direction

 From the eigenvalues and eigenvectors, we calculate the
dielectric constant that in the RPA has the form,
\begin{equation}\label{epsilon}
    \frac{\epsilon(\mathbf{q},\omega)}{\epsilon_0} = 1 -
    \frac{4 \pi e ^2}{\epsilon_0 q ^2 } \, \chi
    (\mathbf{q},\omega)  \, \, \, ,
\end{equation}
being $\chi (\mathbf{q},\omega)$ the susceptibility,
\begin{equation}\label{susceptibility}
\chi(\mathbf{q},\omega)  =   \sum _{i,j,\mathbf{k}}
\frac{n_F(\varepsilon _{i, \mathbf{k}+\mathbf{q}})
-n_F(\varepsilon _{j, \mathbf{k}}) }{\varepsilon _{i,
\mathbf{k}+\mathbf{q}}-\varepsilon _{j, \mathbf{k}}+ \hbar \omega}
f_{(i,\mathbf{k})(j,\mathbf{k}+\mathbf{q})}  \, \, \, \, \, \ ,
\end{equation}
with
\begin{equation}\label{solape}
f_{(i,\mathbf{k})(j,\mathbf{k}')}  =  \left (  \sum _{J,m_J}
(\alpha_{i,\mathbf{k}}^{J,m_J})^* \,\alpha_{j,\mathbf{k}'
}^{J,m_J}  \right ) ^2 \, ,
\end{equation}
where $n_F $ denotes the Fermi-Dirac distribution and $\epsilon
_0$ the dielectric constant of the host semiconductor.
In the GaAs case $\epsilon
_0$=12.5

In Fig.~\ref{fig2} we plot the static susceptibility, $ \chi
(\mathbf{q}) = \chi (\mathbf{q},\omega$=0) as a function of the
wave vector for  $p$=0.02$nm^{-3}$,  and two different
temperatures: $T$=0 where the ground state is ferromagnetic with
$m$=1 and $\xi$=0.777, and $T$=100K$>T_C\sim$57K, where the system
is paramagnetic, $m= \xi$ =0. The {\bf k.p} Hamiltonian has cubic symmetry, and
the susceptibility depends not just on the absolute value of {\bf q},
but also on its orientation. However, for the range of parameters we
are interested in, the dependence is almost negligible. We choose
always the wavector {\bf q} pointing to the [100] direction.
Furthermore, in the ferromagnetic case, since the magnetic
anisotropy is small, the dependence of the susceptibility on the
wave vector orientation is very small and not visible in the scale
of Fig.~\ref{fig2}. The susceptibility depends on $T$ mainly at
small wave vectors, and this dependence occurs because the carrier
screening ability is proportional to the  number of states at the
Fermi surface, and this is smaller in the ferromagnetic phase than
in the paramagnetic phase. At small wave vectors, only the
intraband excitations contribute to the susceptibility and the
static dielectric constant can be written as,
\begin{equation}\label{qtf}
\lim _{q \rightarrow 0} \frac{\epsilon(\mathbf{q})}{\epsilon_0} =
1 +
    \frac{q _{TF} ^2 } {q ^2}
  \, \, \, ,
\end{equation}
and the T dependence of the screening properties can be
characterized by the Thomas Fermi screening length,
$\lambda_{TF}\equiv 1/ q_{TF}$. In Fig.~\ref{fig3} the $T$
dependence of $\lambda_{TF}$ is illustrated. We also show in the
inset of Fig.~\ref{fig3} the $T$-dependence of the ion spin
polarization for the same set of parameters. The screening length
decreases with $T$ and changes near 8$\%$ when going from the
zero $T$ ferromagnetic phase to the paramagnetic phase. Since
this change is due to the variation of the spin polarization, the
Thomas-Fermi length remains constant for $T$ higher than $T_c$.

Finally and in order to compare with experimental information, we
analyze the transport properties of the system. In the
relaxation-time approximation\cite{Jungwirth2} the conductivity
has the form,
\begin{equation}\label{conduc} \sigma_{\alpha,\beta}=
\frac{e^2 }{\hbar V}\sum _{n,\mathbf{k}}
\frac{\tau_{n,\mathbf{k}}}{\hbar} \, \frac{\partial
\varepsilon_{n,\mathbf{k}}}{\partial k _{\alpha}} \,
\frac{\partial \varepsilon_{n,\mathbf{k}}}{\partial k _{\beta}}
\left (- \frac{\partial n_F}{\partial \varepsilon} (\varepsilon
_{n,\mathbf{k}}) \right ) \, \, \, ,
\end{equation}
where $\tau_{n,\mathbf{k}}$ is the elastic scattering time of the
electronic state $(n,\mathbf{k})$, that has the expression,
\begin{eqnarray}\label{tau} \nonumber
\frac{1}{\tau_{n,\mathbf{k}}} & = & \frac{2\pi}{\hbar}\sum _l C_l
\, Q ^2 _l \sum _{n',\mathbf{k}'} \left  | M_ {n,n'}
^{\textbf{k},\textbf{k}'} \right | ^2 \\
& \times &  (1-\cos\theta_{\textbf{k},\textbf{k}'}) \, \delta
(\varepsilon_{n,\mathbf{k}} -\varepsilon_{n',\mathbf{k}'})
\end{eqnarray}
with
\begin{equation}\label{inter}
M_ {n,n'} ^{\textbf{k},\textbf{k}'} =\frac{e ^2 }{\epsilon _0 \,
\,  \epsilon (|\mathbf{k}-\mathbf{k}'|)  \, \,
|\mathbf{k}-\mathbf{k}'|^2 } \, f_{(n,\mathbf{k})(n',\mathbf{k}')}
\, \, \, \, \, \,  ,
\end{equation}
and $C_l$ is the density of  scattering defects with charge $Q_l
\, e$. Using  equations (\ref{tau},\ref{inter}) for the scattering
time we have assumed that the main scattering centers are
Mn$^{2+}$ acceptors, As-antisite defects, and Mn interstitials,
all of them have a Coulomb interaction with the
carriers\cite{Jungwirth2}.  At finite temperature the carriers are
also scattered by short-range spin
fluctuations\cite{fisher,zumsteg}. However we have neglected the
contribution of this spin flip process  to the elastic scattering
time because the Coulomb contribution is orders of magnitude
larger\cite{zumsteg,mjcbreyjav}. The conductivity of DMS at $T$=0,
for different hole densities and different amount of disorder has
been calculated recently in Ref. \cite{Jungwirth2}.
Here we study the dependence of the
conductivity on $T$. We have evaluated expressions
(\ref{conduc}),(\ref{tau}),(\ref{inter}) for different spin
polarization of the holes, which  correspond to different
temperatures. Three factors contribute to the variation of the
electrical conductivity with $T$. i) A change of the thermal Fermi
distribution of the carriers. In the range of temperature of
interest this effect produces only a small variation of the
electronic properties\cite{nota2}. ii) The change of $T$ produces
a change in the dielectric function Fig.\ref{fig2}, and therefore
a change in the matrix element Eq.(\ref{inter}). However this
effect is very weak because in the transport scattering time the
forward scattering is suppressed by the
$(1-\cos\theta_{\textbf{k},\textbf{k}'})$ term and the more
important matrix elements have associated  a large transfer of
momentum for which the electrical susceptibility is practically
$T$ independent, see Fig.\ref{fig2}. iii) The change from the
ferromagnetic to the paramagnetic ground state produces a
reduction of the majority spin Fermi surface area. For that reason
the wave vector difference  that appear in the matrix element
Eq.(\ref{inter}) is bigger in the ferromagnetic phase than in the
paramagnetic  phase, and therefore, the scattering time and the
conductivity are larger in the ferromagnetic phase than in the
paramagnetic phase. This is the main contribution to the variation
of the conductivity with $T$.

\begin{figure}
\includegraphics[clip,width=8cm]{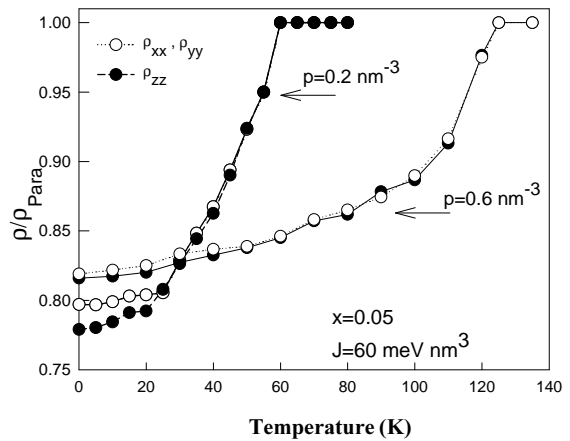} \\  \caption{
Resistivity, normalized to its value in the paramagnetic phase, as
a function of $T$ for Ga$_{1-x}$Mn$_x$As, with $x$=0.05,
$J$=0.06$eVnm^3$ and two different hole densities,
$p$=0.2$nm^{-3}$ and $p$=0.6$nm^{-3}$. } \label{fig1}
\end{figure}

\begin{figure}
\includegraphics[clip,width=8cm]{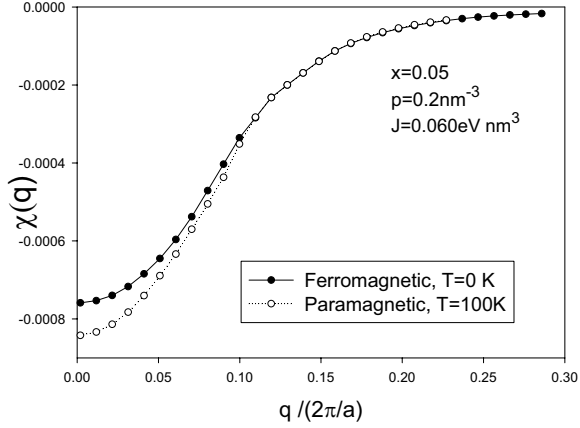} \\  \caption{
Static electrical susceptibility as a function of the wave vector
for a Mn concentration $x$=0.05, a hole density $p$=0.2$nm^{-3}$,
an exchange coupling $J$=0.06$eVnm^3$ and two temperatures: $T$=0
(Ferromagnetic ground state) and $T$=100K (Paramagnetic ground
state). $a$ is the FCC lattice parameter of the host
semiconductor.} \label{fig2}
\end{figure}

\begin{figure}
\includegraphics[clip,width=8cm]{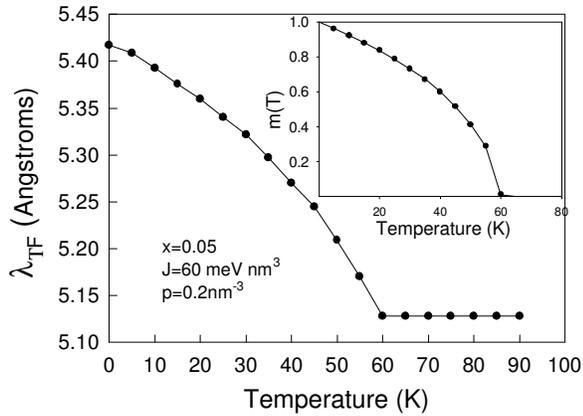} \\  \caption{
Temperature dependence of the Thomas-Fermi length for  a Mn
concentration $x$=0.05, a hole density $p$=0.2$nm^{-3}$, and
exchange coupling $J$=0.06$eVnm^3$. The inset illustrates the T
dependence of the Mn spin polarization.} \label{fig3}
\end{figure}

In Fig.(\ref{fig1}) we plot the resistivity of Ga$_{1-x}$Mn$_x$As,
for $x$=0.05, $J$=60$meVnm^3$ and two different densities,
$p$=0.2$nm^{-3}$ ($T_C=57$K) and $p$=0.6$nm^{-3}$ ($T_C=125$K). We
normalize the resistivity to the value of the resistivity in the
paramagnetic phase. As the temperatures studied are always much
smaller than the Fermi temperature, in the paramagnetic phase the
resistivity is practically $T$ independent. As reported in ref.(\cite{Jungwirth2})
in the ferromagnetic phase there is an
anisotropic effect, and the resistivity is different for
directions parallel and perpendicular to the magnetization. The
anisotropy is $\sim$ 2.5 $\%$ for the $p$=0.2$nm^{-3}$  case and
smaller than $0.5\%$ for the $p$=0.6$nm^{-3}$  case. More
interesting, for both hole densities  there is a bing change, near
$20\%$,  in the value of the resistivity when going from zero $T$
to $T_C$. This change is very similar to the experimentally
observed\cite{Potashnik,Edmonds} in the post growth annealed
samples. Also the overall shape of the resistivity versus $T$
curve is similar to the experimental one.

Therefore  we conclude that our model, that is based in the use of
a)the six band $\mathbf{k} \cdot \mathbf{p}$ model for the band
structure of the DMS, b)the mean field approximation for
describing the thermal fluctuations, c)the RPA for the dielectric
constant  and d)the relaxation-time approximation for the
resistivity,  describe appropriately the electric properties of
DMS and their dependence on $T$. Our theory is able to describe
adequately the experimental dependence of the resistivity on
$T$\cite{Potashnik,Edmonds}.

From our calculations we know that the variation of the
resistivity when going form the ferromagnetic phase to the
paramagnetic phase is due to the different Fermi surfaces  that
have these phases. The wave vector transferred in a scattering
event across the Fermi surface is larger in the ferromagnetic
phase than in the paramagnetic phase and this produces that the
resistivity in the paramagnetic phase is larger than in the
ferromagnetic phase.

We are grateful to G.Platero for helpful discussions. Financial
support is acknowledged from Grants No MAT2002-04429-C03-01 and
MAT2002-04095-C02-01 (MCyT, Spain) and Fundaci\'on Ram\'on Areces.



\begin{thebibliography}{19}
\bibitem[{not()}]{nota2}
\bibinfo{note}{Note that the Fermi temperature is much higher than any other
  temperaure in the system. In particular than the Curie temperature}.
\end{thebibliography}

\end{document}